\documentclass[secnumarabic,preprint,amssymb,aps,pra,nobibnotes,showkeys]{revtex4-1}
\usepackage{docs}
\usepackage[pdftex]{graphicx}
\usepackage{amssymb}
\usepackage{amsmath}
\usepackage{multirow}
\usepackage{bm}
\usepackage{subfigure}
\usepackage{color,soul}
\expandafter\ifx\csname package@font\endcsname\relax\else
 \expandafter\expandafter
 \expandafter\usepackage
 \expandafter\expandafter
 \expandafter{\csname package@font\endcsname}%
\fi
\DeclareRobustCommand\substyle{\name@idx{document substyle}}%
\DeclareRobustCommand\classoption{\name@idx{document class option}}%
\DeclareRobustCommand\classname{\name@idx{document class}}%
\def\name@idx#1#2{%
 {\ttfamily#2}%
 \index{#2\space#1=\string\ttt{#2}\space#1}\index{#1>#2=\string\ttt{#2}}%
}%

\begin{document}
\title{Influence of Magnetic Field and Temperature on Half Width at Half Maximum of Multi-photon Absorption Spectrum in Two-dimensional Graphene}
\author{Cao Thi Vi Ba}
\author{Nguyen Quang Bau}
\email{nguyenquangbau54@gmail.com, nguyenquangbau@hus.edu.vn}
\author{Nguyen Dinh Nam}
\author{Tran Anh Tuan}%
\affiliation{Department of Theoretical Physics, Faculty of Physics, University of Science, Vietnam National University, Hanoi, Vietnam}
\author{Nguyen Thu Huong}
\affiliation{Faculty of Basic Science, Air Defence-Air Force Academy, Kim Son, Son Tay, Hanoi, Vietnam}
\date{\today}
\begin{abstract}
We use the Profile numerical method to calculate the spectral line width, or half width at half maximum (HWHM) of the absorption peaks of multi-photon absorption processes in a two-dimensional graphene system (2DGS) according to important external parameters such as magnetic field and temperature in the presence of strong electromagnetic waves (SEMW). The appearance of these absorption peaks is theoretically obtained from magneto-phonon resonance conditions within the framework of the quantum kinetic equation. The results take into account both scattering mechanisms: electron-optical phonon and electron-acoustic phonon. Under the influence of the magnetic field, according to the increasing photon energy of the SEMW, the graph showing the dependence of the multi-photon nonlinear absorption coefficient on photon energy has the form of absorption spectrum lines following magneto-phonon resonance conditions. When increasing the value of the external magnetic field and the intensity of the SEMW, the intensity of the resonance peaks increases. In addition, the HWHM W of the resonance peaks of multi-photon absorption processes increases with increasing magnetic field $\mathrm{B}$ according to the square root law $\mathrm{W} = \kappa\sqrt{\mathrm{B}}$ but is independent of temperature. The value of the HWHM of the one-photon absorption process is larger than the value of the HWHM of the multi-photon absorption processes. The calculations of the HWHM of the one-photon absorption process in this paper are consistent with previous experimental observations and theoretical calculations. Thus, our calculations of the HWHM of multi-photon absorption processes can serve as reliable predictions for future experiments.
\end{abstract}
\keywords{half width at half maximum, two-dimensional graphene, multi-photon absorption processes, quantum kinetic equation, strong electromagnetic wave, electron-phonon scattering, magneto-phonon resonance}
\maketitle
\section{Introduction}
Since its discovery in 2004 \cite{akgeim, ksno}, Graphene has quickly risen to the top of the list of new-generation semiconductor materials that a large number of physicists are interested in researching, both theoretically and experimentally. Unlike traditional two-dimensional semiconductor materials where electrons are Schrodinger Fermions, electrons in Graphene behave as massless Dirac Fermions described by the Weyl equation. More specifically, when the Graphene sheet is placed under the influence of an external magnetic field, the electronic energy spectrum depends on the magnetic field B and the Landau index $\mathrm{n}$ according to the square root law \cite{ksno}. This leads to the spacing between electron Landau energy levels becoming unequally spaced, contrary to the properties of low-dimensional semiconductor systems  \cite{eps,eps1}. The difference between the electronic energy spectrum of graphene and conventional two-dimensional semiconductors causes Graphene to have many unique and novel optical and electrical properties, especially when the system is placed in a strong electromagnetic wave (SEMW). 

The problem of nonlinear absorption of strong electromagnetic waves by a semiconductor material system, when the system is placed under the influence of an external magnetic field, has long been of interest to research by physicists. In low-dimensional semiconductor systems and Graphene, this problem can be solved by many different methods, such as the operator projection method \cite{bau2, bdh}, perturbation theory method \cite{lvtung, pham, phuc}, effective mass approximation \cite{hjap}, Boltzmann kinetic equation \cite{kry} and quantum kinetic equation method \cite{bau1, tuan}. In these works, the authors have made theoretical calculations and given explicit analytical expressions for the absorption coefficient or nonlinear absorption power of strong electromagnetic waves in traditional low-dimensional semiconductor systems and Graphene. The results show the dependence of the absorption coefficient or absorption power on the photon energy and intensity of the SEMW, external magnetic field, and temperature of the system. In addition, the authors also calculated the dependence of the absorption spectral line width on external parameters of the system such as magnetic field and temperature. However, most of these studies only focus on calculating the absorption spectral line width of the one-photon absorption (1PA) process; multi-photon absorption processes have not been specifically mentioned. 

The absorption spectral line width also known as The Full Width at Half maximum (FWHM) refers to the disparity between two distinct values of the dependent variable (usually the frequency or photon energy of SEMW), wherein the absorption coefficient attains a magnitude equivalent to half of its maximum value. In low-dimensional semiconductor systems, such as two-dimensional quantum Wells with different potential profiles, FWHM has been calculated numerically as functions depending on the external field, the temperature, well width, alloy composition, and doping position both theoretically \cite{lvtung, pham, unuma} and experimentally \cite{tndr1,tndr2}. These results show that FWHM increases with external magnetic field B and temperature T but decreases with Well width. In particular, theoretically, the dependence of FWHM on magnetic field B and temperature T obeys the square root law \cite{van2, lvtung}. In Graphene, considering the electron-optical phonon scattering mechanism, previous theoretical calculations of the half-width at half-maximum (HWHM) (note that HWHM is half of FWHM) of 1PA have also shown a square-root dependence on the external magnetic field and is almost independent of temperature \cite{tuan, bdh,phuc, orlita,ji}. Notwithstanding the significant outcomes of the aforementioned studies, there is still a lack of a systematic study of the influence of magnetic field and temperature on HWHM in Graphene, especially HWHM calculations of the multi-photon absorption processes. 

In this study, based on the quantum kinetic equation method for electrons in graphene performed in previous work \cite{tuan}, we perform a detailed and systematic calculation of the dependence of HWHM of the multi-photon absorption process on the magnetic field and temperature. Also, different from previous studies, we include HWHM calculations in both electron-optical phonon and electron-acoustic phonon scattering mechanisms. Our calculations are compared to the recent theoretical and experimental results, which show the difference and the novelty of the results. The paper is organized as follows: In Sec. II, we will present in detail the method of calculating the HWHM width based on the explicit analytical expression of the Multi-photon Nonlinear Absorption Coefficient (MPNAC) in both scattering mechanisms. In Sec. III, we perform detailed numerical calculations and plot the dependence of MPNAC on photon energy and intensity of the SMEW. Then, using the Profile numerical calculation method \cite{pro}, we calculate the influence of magnetic field and temperature on HWHM and provide physical discussions. Finally, conclusions are given in Sec. IV.  

\section{Theoretical Calculation} 
In this paper, we examine a two-dimensional Graphene system (2DGS) in which electrons move freely in the (x-y) plane and a uniform static magnetic field of strength B is exerted in the z-direction $\mathbf{B}=(0,0, B)$. The wave function and the corresponding energy are written as \cite{ando}  
\begin{align}
\left| {{\rm{n}},{{\bf{k}}_y}} \right\rangle  = \frac{{{C_{\rm{n}}}}}{{\sqrt {{L_y}} }}\exp \left( {i{k_y}y} \right)\left[ {\begin{array}{*{20}{c}}
{{S_{\rm{n}}}{\Phi _{\left| {\rm{n}} \right| - 1}}\left( {x - {k_y}\ell _B^2} \right)}\\
{{\Phi _{\left| {\rm{n}} \right|}}\left( {x - {k_y}\ell _B^2} \right)}
\end{array}} \right]
\end{align}
\begin{align}
    {{\mathcal{E}} _{\rm{n}}} = {S_{\rm{n}}}\hbar {\omega _B}\sqrt {\left| {\rm{n}} \right|} 
\end{align}
Here, $\mathrm{n}$ being the Landau indices, ${C_{\rm{n}}} = {1 \mathord{\left/
 {\vphantom {1 {\sqrt 2 }}} \right.
 \kern-\nulldelimiterspace} {\sqrt 2 }}$ when ${\rm{n}} \ne 0$, and ${C_{\rm{n}}} = 1$ when ${\rm{n}} = 0$. $L_y$ is the normalization length in the y direction of the system, and $k_y$ is the wave vector along the y-axis. ${S_{\rm{n}}} = 1$ and $-1$ represent the conduction and valence bands, respectively. ${\Phi _{\left| {\rm{n}} \right|}}\left( x \right) = {{{i^{\left| {\rm{n}} \right|}}\exp \left( { - {{{x^2}} \mathord{\left/
 {\vphantom {{{x^2}} {2\ell _B^2}}} \right.
 \kern-\nulldelimiterspace} {2\ell _B^2}}} \right){{\rm H}_{\left| {\rm{n}} \right|}}\left( {{x \mathord{\left/
 {\vphantom {x {{\ell _B}}}} \right.
 \kern-\nulldelimiterspace} {{\ell _B}}}} \right)} \mathord{\left/
 {\vphantom {{{i^{\left| {\rm{n}} \right|}}\exp \left( { - {{{x^2}} \mathord{\left/
 {\vphantom {{{x^2}} {2\ell _B^2}}} \right.
 \kern-\nulldelimiterspace} {2\ell _B^2}}} \right){{\rm H}_{\left| {\rm{n}} \right|}}\left( {{x \mathord{\left/
 {\vphantom {x {{\ell _B}}}} \right.
 \kern-\nulldelimiterspace} {{\ell _B}}}} \right)} {\sqrt {{2^{\left| {\rm{n}} \right|}}\left| {\rm{n}} \right|!\sqrt \pi  {\ell _B}} }}} \right.
 \kern-\nulldelimiterspace} {\sqrt {{2^{\left| {\rm{n}} \right|}}\left| {\rm{n}} \right|!\sqrt \pi  {\ell _B}} }}$ is the normalized harmonic oscillator function, in which, ${\ell _B} = \sqrt {{\hbar  \mathord{\left/
 {\vphantom {\hbar  {eB}}} \right.
 \kern-\nulldelimiterspace} {eB}}} $ is the magnetic length, ${{\rm H}_{\rm{n}}}\left( x \right)$ is the n-th order Hermite polynomial, and $\hbar {\omega _B} = {{\sqrt 2 \gamma } \mathord{\left/
 {\vphantom {{\sqrt 2 \gamma } {{\ell _B}}}} \right.
 \kern-\nulldelimiterspace} {{\ell _B}}}$ is the effective
magnetic energy with $\gamma = 6.46{\text{eV}}{\text{.}}\mathop {\text{A}}\limits^{\text{o}}$  is the band parameter. 

\begin{figure}[!htb]
\centering
\includegraphics[scale = 0.45]{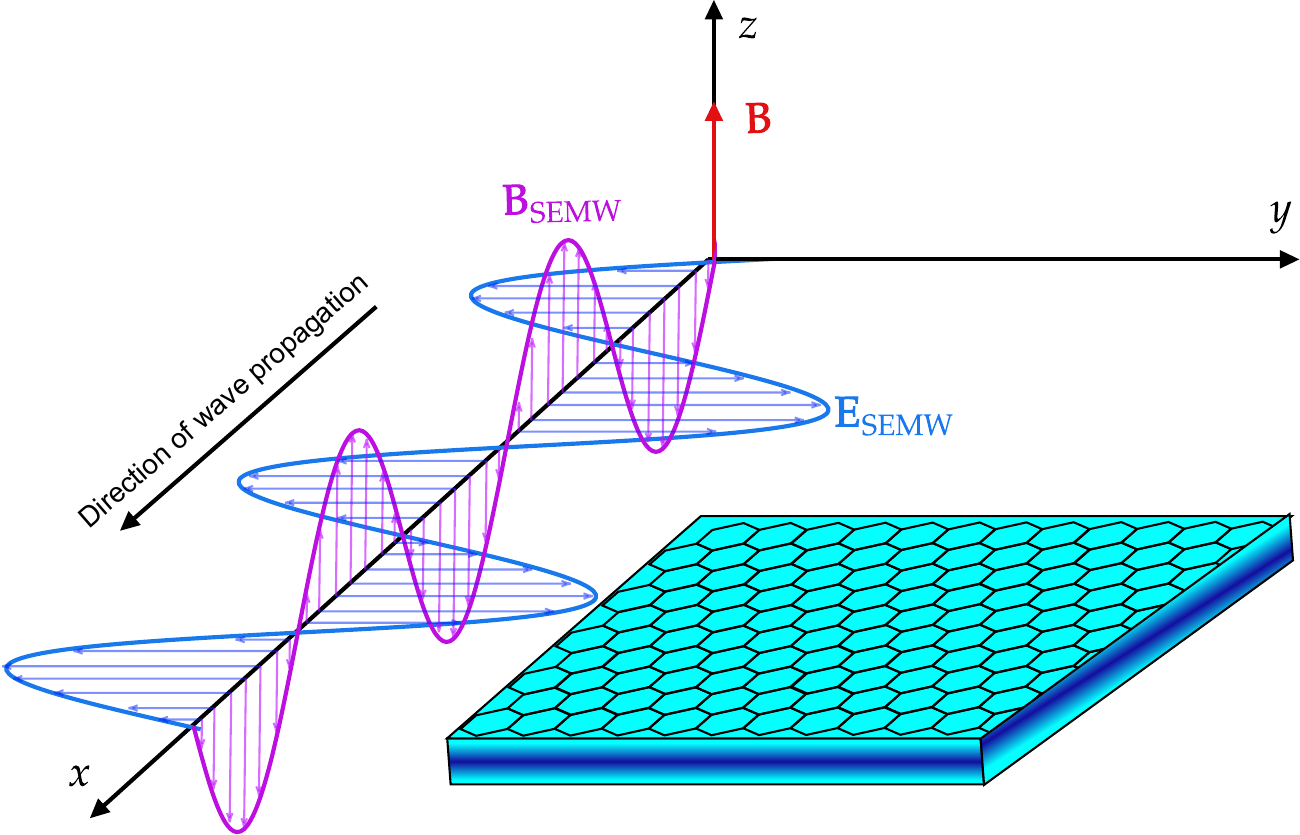}
  \caption{(Color online) Illustration of SEMW propagation in a Graphene sheet under the influence of a magnetic field perpendicular to the z direction.}
  \label{lantruyen}
\end{figure}

If the 2DGS is placed under the influence of a SEMW propagating along the x-axis with the electric field vector ${\mathbf{E}_{\rm{SEMW}}} = \left( {0,{{\rm{E}}_0}\sin \Omega t,0} \right)$ (${{\rm{E}}_0}$ and $\Omega$ are the intensity and frequency of the SEMW) (see Fig. \ref{lantruyen}), within the framework of the quantum kinetic equation method, using the calculations performed previously \cite{tuan}, we obtain the explicit analytical expressions of MPNAC in both cases of electron-optical phonon scattering and electron-acoustic phonon scattering

\begin{align}\label{op}
\alpha _{{\rm{op}}}^B = \frac{{4\pi {n_0}{\rm{D}}_{{\rm{op}}}^2\hbar \Omega \overline {{{\cal{N}}_{0}}} }}{{c\sqrt {{\kappa _\infty }} {\rm{E}}_0^2\rho \ell _B^2{\omega _0}}}\sum\limits_{{{\rm{n}}^\prime },{\rm{n}}} {\overline {{{\mathcal{F}}_{\rm{n}}}} \sum\limits_\ell  {{{\cal A}_\ell }\delta \left( {{{\cal E}_{{{\rm{n}}^\prime }}} - {{\cal E}_{\rm{n}}} + \hbar {\omega _0} - \ell \hbar \Omega } \right)} } 
\end{align}
\begin{align}\label{ac}
    \alpha _{{\rm{ac}}}^B = \frac{{4\pi {n_0}{\rm{D}}_{{\rm{ac}}}^{\rm{2}}\Omega {k_B}T}}{{c\sqrt {{\kappa _\infty }} {\rm{E}}_0^2\rho \ell _B^2\nu _{\rm{s}}^2}}\sum\limits_{{{\rm{n}}^\prime },{\rm{n}}} {\overline {{{\mathcal{F}}_{\rm{n}}}} \sum\limits_\ell  {{{\cal A}_\ell }\delta \left( {{{\cal E}_{{{\rm{n}}^\prime }}} - {{\cal E}_{\rm{n}}} - \ell \hbar \Omega } \right)} } 
\end{align}
where, $n_0$ is electron density in Graphene, $\rm{D_{op}}$ and $\rm{D_{ac}}$ are the deformation potential coupling constant of optical phonon and acoustic phonon, respectively. c is the speed of light in a vacuum, $\kappa _\infty$ is the high-frequency dielectric constant in graphene. $\rho$ and $\nu_{\rm{s}}$ are the mass density and the sound velocity in graphene. $k_B$ is the Boltzmann constant, and T is the temperature of 2DGS.  $\overline {{{\cal{N}}_0}}  = {\left[ {\exp \left( {{{\hbar {\omega _0}} \mathord{\left/
 {\vphantom {{\hbar {\omega _0}} {{k_B}T}}} \right.
 \kern-\nulldelimiterspace} {{k_B}T}}} \right) - 1} \right]^{ - 1}}$ is the Bose-Einstein distribution function for optical phonons and $\overline {{\mathcal{F}_{\text{n}}}}  = {\left\{ {1 + \exp \left[ {{{\left( {{\mathcal{E}_{\text{n}}} - {\mathcal{E}_{\text{F}}}} \right)} \mathord{\left/
 {\vphantom {{\left( {{\mathcal{E}_{\text{n}}} - {\mathcal{E}_{\text{F}}}} \right)} {\left( {{k_B}T} \right)}}} \right.
 \kern-\nulldelimiterspace} {\left( {{k_B}T} \right)}}} \right]} \right\}^{ - 1}}$ is the Fermi-Dirac distribution function for electrons with $ {{\cal E}_{\rm{F}}}$ is Fermi energy. ${\mathcal{A}}_\ell$ is the dimensionless parameter characterizing $\ell$-photon absorption process from Eqs. (27), (28), and (29) in Ref. \cite{tuan}. 

 The quantum kinetic equation method used in this paper allows for an investigation of the quantum theory involving the physical properties of many-body systems, including semiconductor systems, advanced material systems, and low-dimensional systems, i.e., especially typical two-dimensional Dirac material systems such as Graphene. However, the analytical expressions shown here cannot be applied to general patterns such as semiconductor systems, advanced material systems, and low-dimensional systems, i.e. Obviously, the state of electrons in different many-body systems is different, leading to differences in the wave function and energy spectrum of the electron as well as the interaction behavior of the electron with other quasiparticles in the system. This is a simple theoretical basis for showing that the expressions of MPNAC in different systems are different.

\begin{figure}[!htb]
\centering
\includegraphics[scale = 0.55]{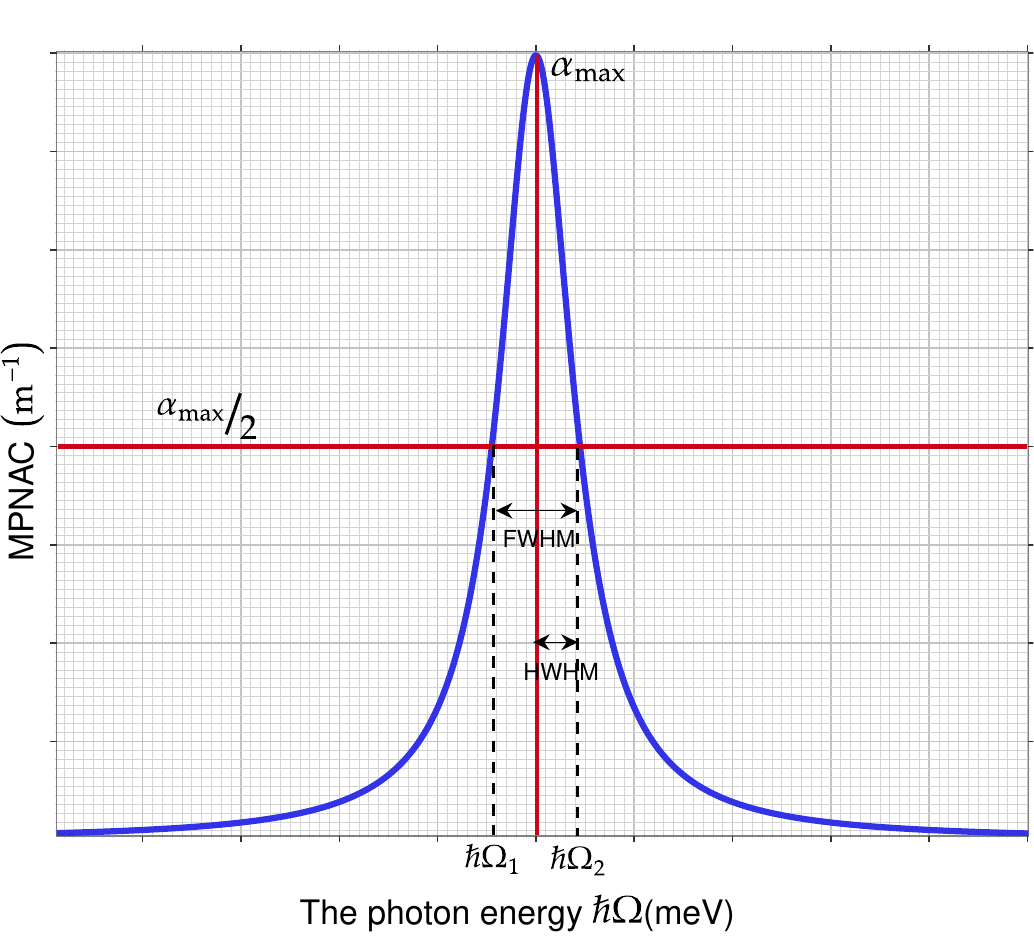}
  \caption{(Color online) Illustration of the Profile method to find FWHM and HWHM from the graph showing the dependence of MPNAC on the photon energy of the SEMW.}
  \label{profile}
\end{figure} 

The Dirac Delta function in Eqs. \eqref{op} and \eqref{ac} will be replaced by Lorentzians of width ${\Gamma _{{{\rm{n}}^\prime },{\rm{n}}}}$ \cite{van1,van2,van3}
\begin{align}
    \delta \left( {{{\cal E}_{{{\rm{n}}^\prime }}} - {{\cal E}_{\rm{n}}} + \hbar {\omega _{\bf{q}}} - \ell \hbar \Omega } \right) = \frac{1}{\pi} \frac{{{\Gamma _{{{\rm{n}}^\prime },{\rm{n}}}}}}{{{{\left( {{{\cal E}_{{{\rm{n}}^\prime }}} - {{\cal E}_{\rm{n}}} + \hbar {\omega _{\bf{q}}} - \ell \hbar \Omega } \right)}^2} + \Gamma _{{{\rm{n}}^\prime },{\rm{n}}}^2}}
\end{align}
with $\Gamma _{{{\text{n}}^\prime },{\text{n}}}^2 =\displaystyle\sum\limits_{\mathbf{q}} {\left( {\overline {{\mathcal{N}_{\mathbf{q}}}}  + \dfrac{1}{2} \pm \frac{1}{2}} \right){{\left| {{{\text{M}}_{{{\text{n}}^\prime },{\text{n}}}}\left( {\mathbf{q}} \right)} \right|}^2}}  $, in which ${{{\rm M}_{{{\rm{n}}^\prime },{\rm{n}}}}\left( {\bf{q}} \right)}$ is the electron-phonon scattering matrix factor in Eq. (22) from Ref. \cite{tuan}. $\hbar\omega_{\bf{q}}$ is the phonon energy, $\hbar {\omega _{\bf{q}}} \approx \hbar {\omega _0} = 162{\rm{meV}}$ for the case of non-dispersive optical phonons \cite{ando} and $\hbar {\omega _{\bf{q}}} = \hbar {\nu _s}{\rm{q}}$ (q being the wave vector of phonon) for acoustic phonons, the acoustic phonon energy is considered small and ignored in the Dirac Delta function \cite{van1}.

We now present the method for calculating FWHM and HWHM (see Fig. \ref{profile}). From the graph showing the dependence of MPNAC on the photon energy of the SEMW, for any absorption peak $\alpha_{max}$, we solve the equation $\alpha \left( {\hbar \Omega } \right) = {{{\alpha _{{\rm{max}}}}} \mathord{\left/
 {\vphantom {{{\alpha _{{\rm{max}}}}} 2}} \right.
 \kern-\nulldelimiterspace} 2}$. The results are two solutions $\hbar\Omega_1$ and $\hbar\Omega_2$. From there we can determine ${\mathrm{FWHM}} = \hbar\Omega_2 - \hbar\Omega_1 $, and ${\mathrm{HWHM}} = {\rm{W}} = {{\left( {\hbar {\Omega _2} - \hbar {\Omega _1}} \right)} \mathord{\left/
 {\vphantom {{\left( {\hbar {\Omega _2} - \hbar {\Omega _1}} \right)} 2}} \right.
 \kern-\nulldelimiterspace} 2} $. In the next section, we will perform detailed numerical calculations of MPNAC and HWHM(W) with the aid of computer programs to elucidate important physical properties. 
\section{Numerical analysis and discussions}
In this section, we present a comprehensive analysis of the numerical assessment of MPNAC and HWHM under the influence of an external magnetic field, focusing on two distinct electron-phonon scattering mechanisms. The parameters used for calculations are provided in \cite{tuan}. In the numerical analysis of this paper, we only consider Landau levels from $\mathrm{n} = 0$ state to ${\mathrm{n}}^\prime = 1$ state, so that we can clearly examine the HWHM of each absorption peak of multi-photon absorption processes. 

\begin{figure}[!htb]
\centering
\subfigure[][Electron-optical phonon scattering. Here, ${\mathrm{T}} = 4.2 {\mathrm{K}}, {\mathrm{E}}_{0} = 1.5 \times 10^{7} {\mathrm{V/m}}$.]{\label{3a}\includegraphics[width=0.47\linewidth]{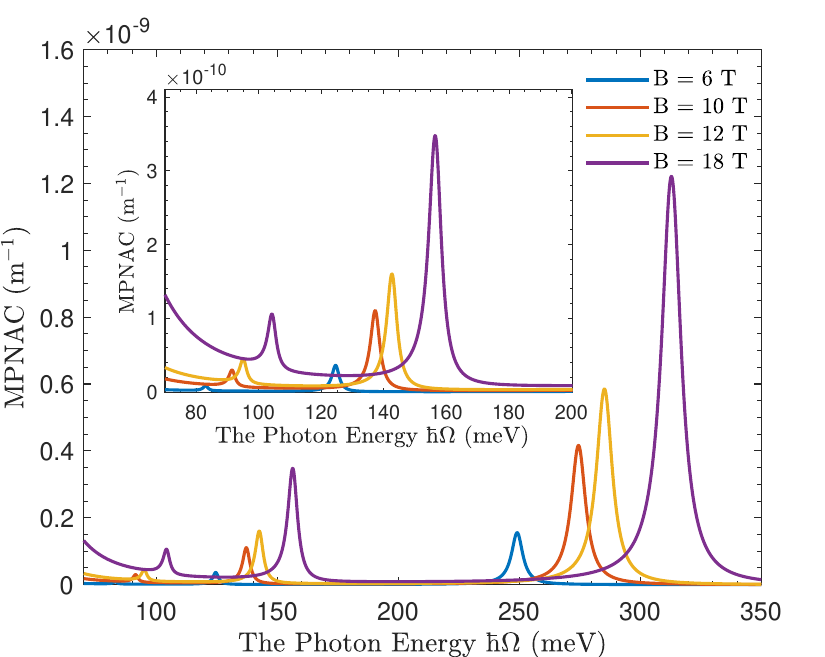}}
\subfigure[][Electron-acoustic phonon scattering. Here, ${\mathrm{T}} = 4.2 {\mathrm{K}}, {\mathrm{E}}_{0} = 1.0 \times 10^{6} {\mathrm{V/m}}$.]{\label{3b}\includegraphics[width=0.50\linewidth]{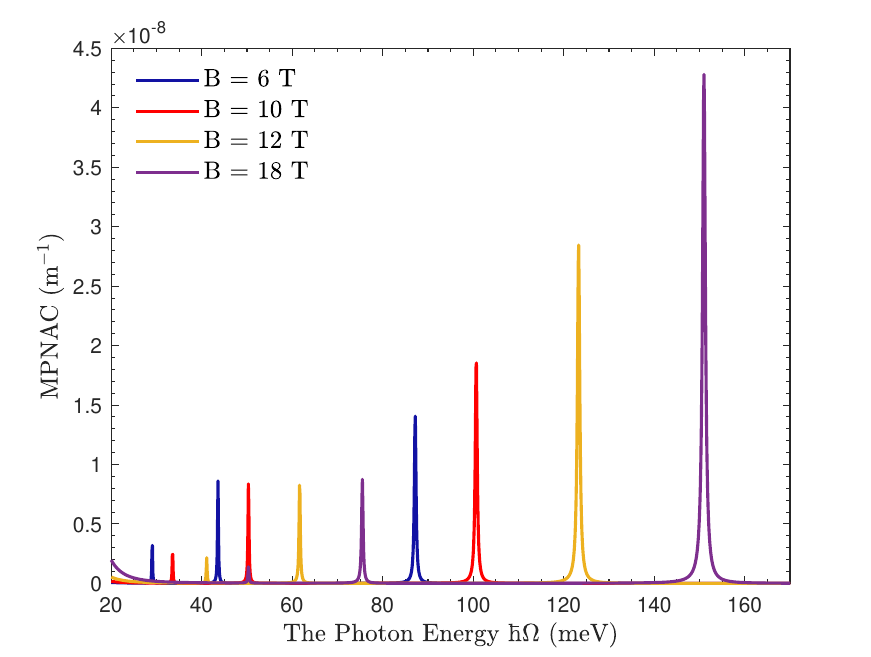}}
\caption{(Color online) The dependence of MPNAC on the photon energy with the different values of the external magnetic field.}
\label{3}
\end{figure}

In Figs. \ref{3a} and \ref{3b}, we show our calculations for MPNAC as a function of the photon energy of the SEMW in both two cases of the electron scattering mechanism. We observe that, for each value of the external magnetic field B (or the intensity $\rm{E}_0$), three absorption spectrum lines appear corresponding to the one-, two-, and three-photon absorption processes. As in previous works \cite{tuan, bdh}, the positions of the spectral lines or the resonant peaks follow the Magneto-phonon resonance condition (MPRC) $\ell \hbar \Omega  = \hbar {\omega _B}\left( {\sqrt {\left| {{{\rm{n}}^\prime }} \right|}  - \sqrt {\left| {\rm{n}} \right|} } \right) + \hbar {\omega _0}$ (for the electron-optical phonon scattering mechanism) and $\ell \hbar \Omega  = \hbar {\omega _B}\left( {\sqrt {\left| {{{\rm{n}}^\prime }} \right|}  - \sqrt {\left| {\rm{n}} \right|} } \right)$ (for the electron-acoustic phonon scattering mechanism) with $\ell = 1, 2$, and 3. These theoretical results are found when the argument of the Dirac Delta function in Eqs. \eqref{op} and \eqref{ac} is equal to 0. Physically, these relations are a consequence of the law of conservation of energy. The positions of the resonance peaks in both electron-acoustic phonon and electron-optical phonon scattering mechanisms are given in Tab. \ref{tab1} below. 

\begin{table}[!htb]
\centering
\caption{The positions of the resonance peaks in both cases of electron scattering under the influence of an external magnetic field are found using numerical methods in Fig. \ref{3}.}
\begin{ruledtabular}
\begin{tabular}{c ccc || ccc}
\multirow{2}{*}{\textbf{\begin{tabular}[c]{@{}c@{}}The positions\\ of peaks\end{tabular}}} &
  \multicolumn{3}{c||}{\textbf{Electron-optical phonon scattering}} &
  \multicolumn{3}{c}{\textbf{Electron-acoustic phonon scattering}} \\ 
        & \multicolumn{1}{c}{1PA (meV)} & \multicolumn{1}{c}{2PA (meV)} & 3PA (meV) & \multicolumn{1}{c}{1PA (meV)} & \multicolumn{1}{c}{2PA (meV)} & 3PA (meV) \\ \hline
B = 6T  & \multicolumn{1}{c}{249.3}     & \multicolumn{1}{c}{124.5}     & 83.0      & \multicolumn{1}{c}{87.2}      & \multicolumn{1}{c}{43.6}      & 29.1      \\
B = 10T & \multicolumn{1}{c}{274.3}     & \multicolumn{1}{c}{137.1}     & 91.4      & \multicolumn{1}{c}{112.5}     & \multicolumn{1}{c}{56.3}      & 37.5      \\
B = 12T & \multicolumn{1}{c}{285.0}     & \multicolumn{1}{c}{142.7}     & 95.0      & \multicolumn{1}{c}{123.3}     & \multicolumn{1}{c}{61.6}      & 41.1      \\
B = 18T & \multicolumn{1}{c}{312.9}     & \multicolumn{1}{c}{156.2}     & 104.3     & \multicolumn{1}{c}{151.4}     & \multicolumn{1}{c}{75.5}      & 50.3      \\ 
\end{tabular}
\end{ruledtabular}
\label{tab1}
\end{table}

The results in Figs. \ref{3a}, \ref{3b}, and Tab. \ref{tab1} show that as the magnetic field increases, the position of the resonance peaks shifts to the right and the intensity of the resonance peaks also increases. This can be explained from MPRC, when the magnetic field increases, the effective magnetic energy $\hbar\omega_B$ also increases (recall, $\hbar {\omega _B} = {{\sqrt 2 \gamma } \mathord{\left/
 {\vphantom {{\sqrt 2 \gamma } {{\ell _B}}}} \right.
 \kern-\nulldelimiterspace} {{\ell _B}}} = \gamma \sqrt {{{2eB} \mathord{\left/
 {\vphantom {{2eB} \hbar }} \right.
 \kern-\nulldelimiterspace} \hbar }} $) leading to $\hbar\Omega$ increasing, the resonance peaks will shift to the right (with larger $\hbar\Omega$ values). In addition, the appearance of an external magnetic field causes a harmonic confinement potential, thereby increasing the probability of electron-phonon scattering, leading to an increase in MPNAC. 
 
\begin{figure}[!htb]
\centering
\subfigure[][Electron-optical phonon scattering]{\label{4a}\includegraphics[width=0.49\linewidth]{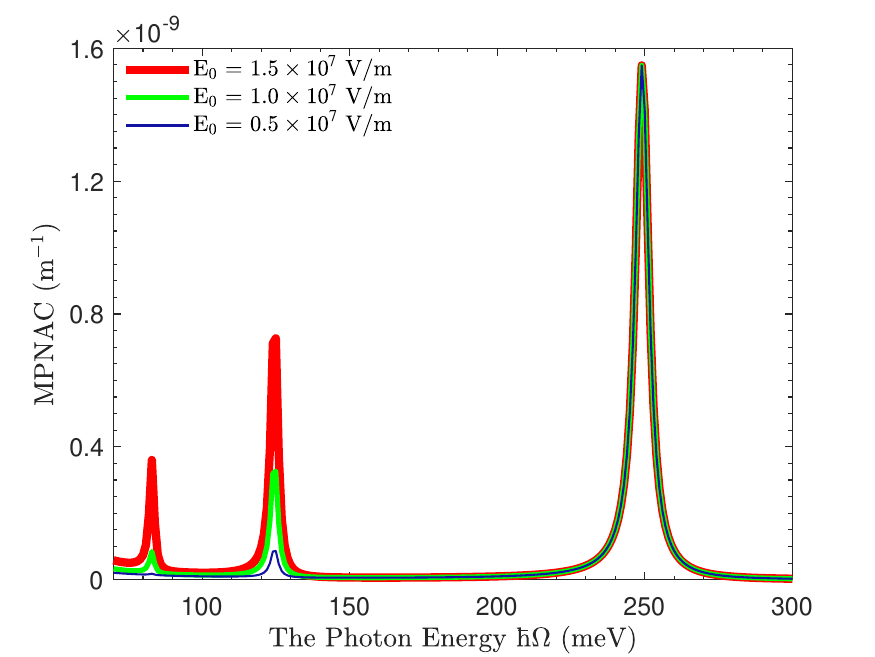}}
\subfigure[][Electron-acoustic phonon scattering]{\label{4b}\includegraphics[width=0.49\linewidth]{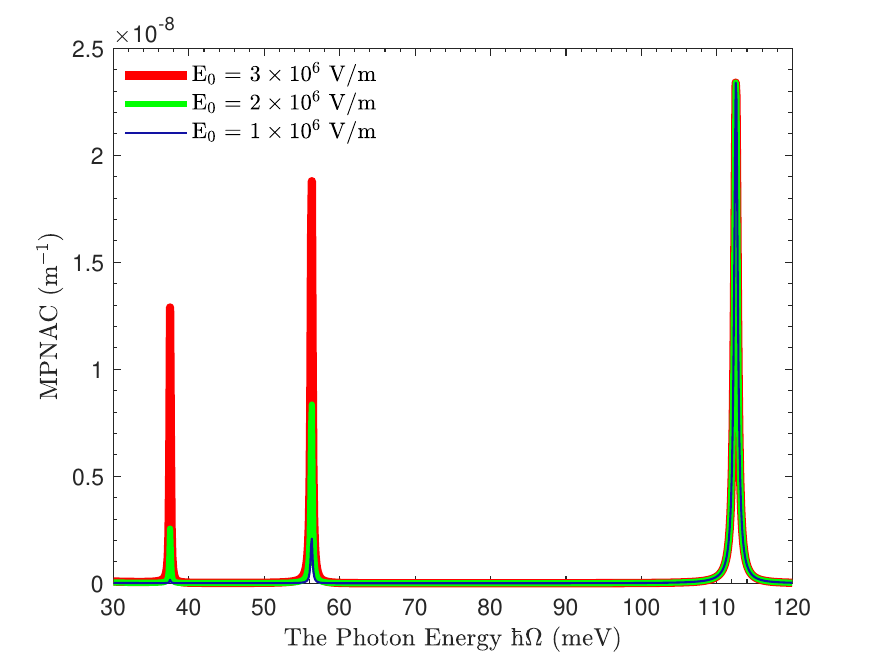}}
\caption{(Color online) The dependence of MPNAC on the photon energy with the different values of the intensity of the SEMW. Here, $\mathrm{T = 4.2 K}$, $\mathrm{B} = 6 \mathrm{T}$. }
\label{4}
\end{figure}

Next, we pay attention to the influence of the intensity $\rm{E}_0$ of SEMW on the absorption spectrum. Because in MPRC there is no appearance of $\rm{E}_0$, it can be seen that $\rm{E}_0$ does not affect the position of the resonance peaks but only changes their intensity, this is shown in Figs. \ref{4a} and \ref{4b}. Indeed, considering the electron-optical phonon scattering mechanism, with B = 6T, according to MPRC and Tab. \ref{tab1}, we can determine the positions of the resonance peaks in Fig. \ref{4a}, specifically $\hbar {\Omega _1} \approx 249.3{\rm{meV}}, \hbar {\Omega _2} \approx 124.5{\rm{meV}}$ and $\hbar {\Omega _3} \approx 83{\rm{meV}}$ corresponding to the absorption peaks of one, two and three photons. The results for the one- and two-photon absorption (2PA) peak positions were found in \cite{phuc}, the three-photon absorption (3PA) peak position result is our new finding. Besides, when $\rm{E}_0 = 0.5 \times 10^7 \mathrm{V/m}$, we only observe the appearance of one- and two-photon absorption peaks, in which the intensity of the 2PA peak is $5.6\%$ of the intensity of the 1PA peak. However, when the intensity of $\rm{E}_0$ increases, with $\rm{E}_0 = 1.5 \times 10^7 \mathrm{V/m}$, the 3PA peak appears and its intensity equals to $23.3\%$ of the intensity of the 1PA peak. This result shows that, in the presence of a strong electromagnetic wave, the effect of the 3PA process is significant. In future experiments, this could be seen as an important criterion that will help to find the location of multi-photon absorption spectral lines. In addition, it can be seen that the 1PA peak intensity has a magnitude that does not depend on $\rm{E}_0$, simply because this is a linear absorption process. 

\begin{figure}[!htb]
\centering
\subfigure[][Electron-optical phonon scattering. Here, ${\mathrm{T}} = 4.2 {\mathrm{K}}, {\mathrm{E}}_{0} = 1.0 \times 10^{7} {\mathrm{V/m}}$.]{\label{5a}\includegraphics[width=0.51\linewidth]{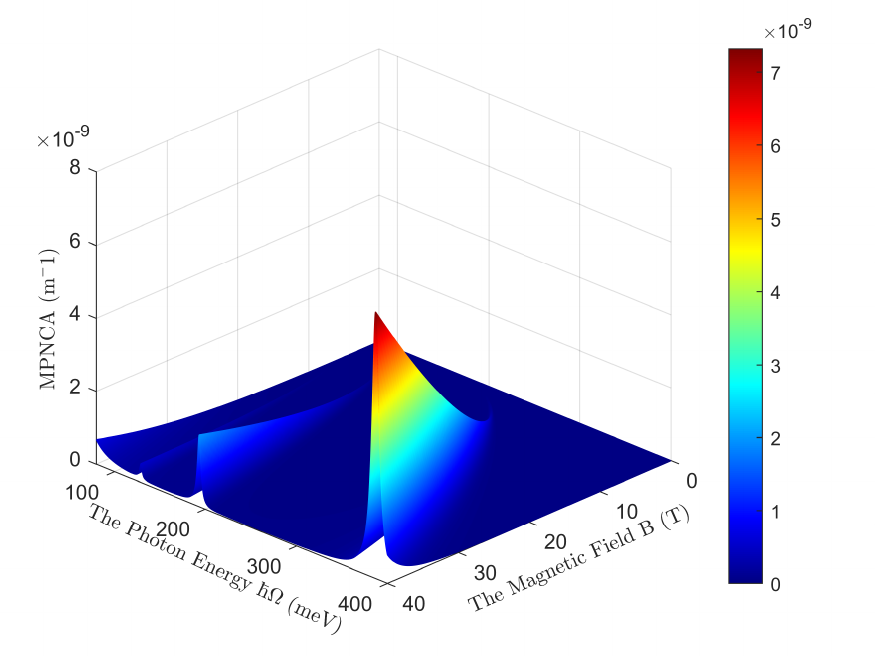}}
\subfigure[][Electron-acoustic phonon scattering. Here, ${\mathrm{T}} = 4.2 {\mathrm{K}}, {\mathrm{E}}_{0} = 1.8 \times 10^{6} {\mathrm{V/m}}$.]{\label{5b}\includegraphics[width=0.47\linewidth]{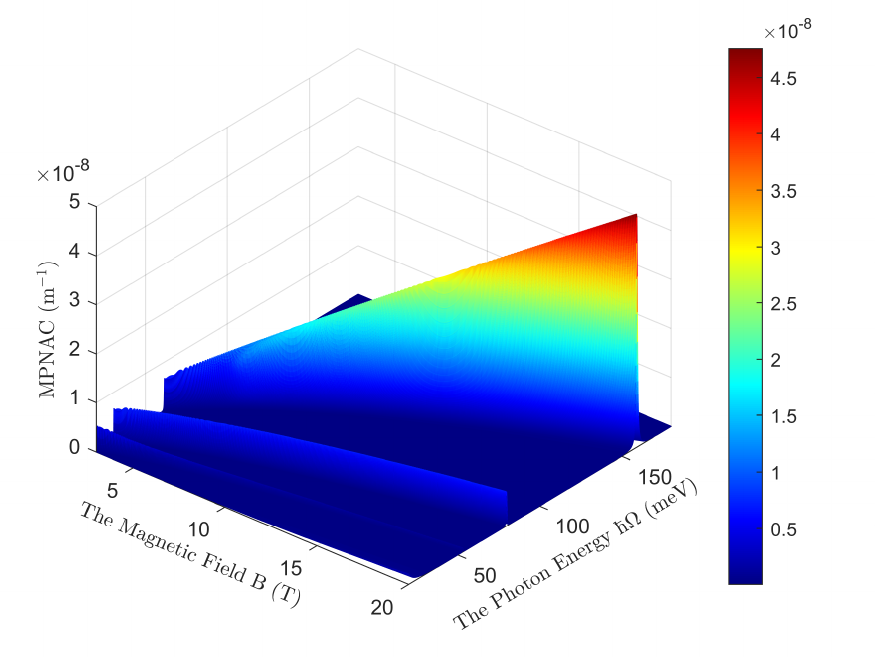}}
\caption{(Color online) MPNAC as a function of the magnetic field and the photon energy with the side view.}
\label{5}
\end{figure}

\begin{figure}[!htb]
\centering
\subfigure[][Electron-optical phonon scattering. Here, ${\mathrm{T}} = 4.2 {\mathrm{K}}, {\mathrm{E}}_{0} = 1.0 \times 10^{7} {\mathrm{V/m}}$.]{\label{6a}\includegraphics[width=0.47\linewidth]{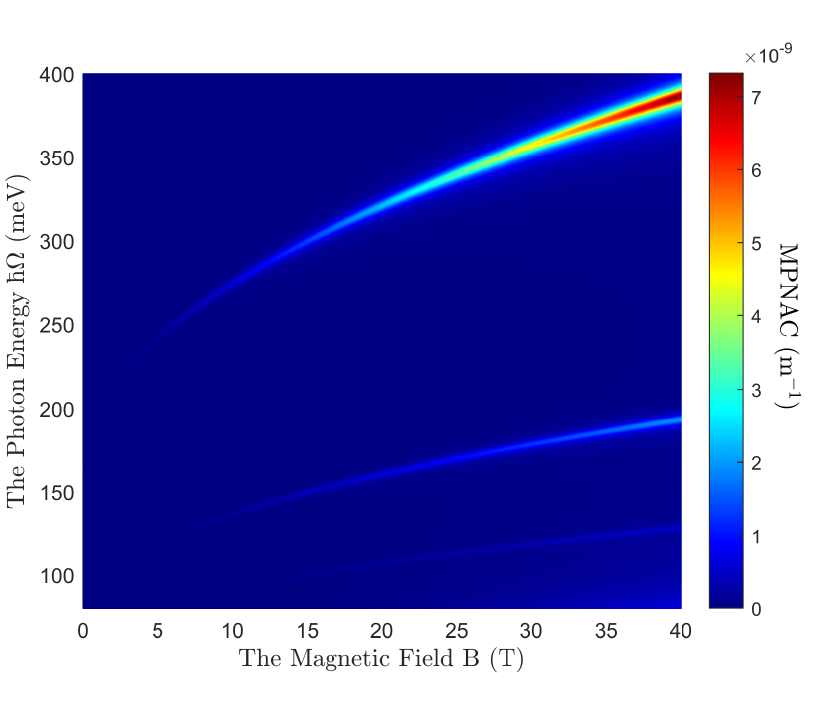}}
\subfigure[][Electron-acoustic phonon scattering. Here, ${\mathrm{T}} = 4.2 {\mathrm{K}}, {\mathrm{E}}_{0} = 1.8 \times 10^{6} {\mathrm{V/m}}$.]{\label{6b}\includegraphics[width=0.50\linewidth]{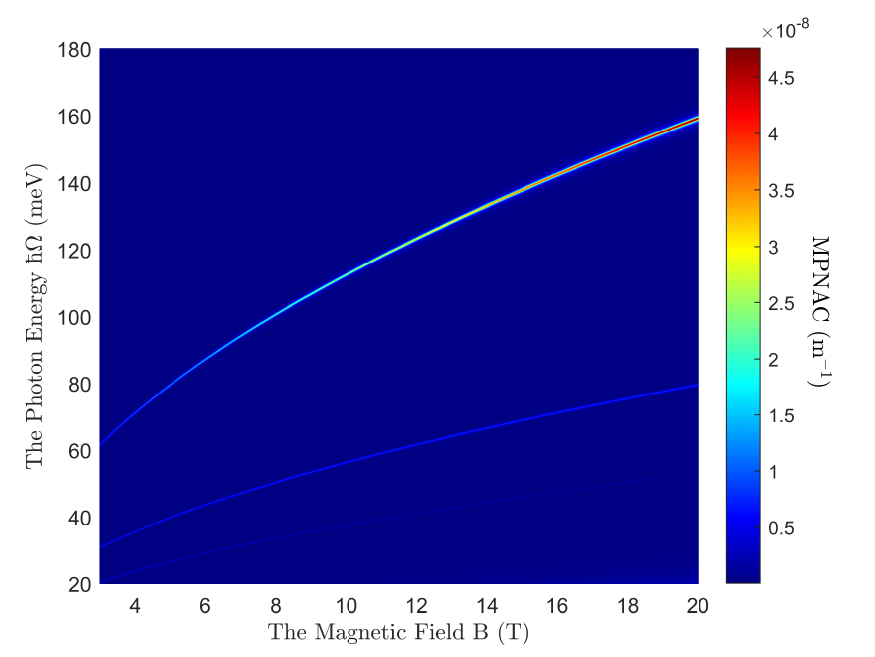}}
\caption{(Color online) The density graph illustrates the relationship between the MPNAC and the magnetic field and the photon energy of SEMW.}
\label{6}
\end{figure}

To further clarify the influence of the magnetic field on the absorption spectrum of electrons in 2DGS, we present a three-dimensional graph showing the relationship between MPNAC, the location of the resonance peaks, and the value of the external magnetic field in Figs. \ref{5} and \ref{6}. The appearance of the resonance peaks in Figs. \ref{5a} and \ref{5b} can be explained by the MPRC as presented above. For a more general view, we provide a graph representing the density of MPNAC according to the magnetic field and photon energy, as shown in Figs. \ref{6a} and \ref{6b}. We can deduce that the photon energy at the resonant peak locations appears to depend linearly on the square root of the magnetic field. This can be generalized to be true for all resonance peaks of multi-photon absorption processes. This deduction is consistent with previous theoretical calculations by Kubo Formula \cite{yang}, by the projection operator method \cite{bdh}, and the experimental observations \cite{ji} for the 1PA process. Also from Figs. \ref{6a} and \ref{6b}, we see that the FWHM of the resonance peak in all three absorption processes broadens as the magnetic field increases. In particular, for the electron-acoustic phonon scattering mechanism, this spectral line broadening is relatively small and difficult to observe. We will clarify this in the HWHM calculations below 

Plotted in Figs. \ref{7a} and \ref{7b} is the HWHM W as a function of the magnetic field $\mathrm{B}$. According to our calculations, the dependence of HWHM on the magnetic field follows the square root law. First of all, we consider the electron-optical phonon scattering mechanism, because this is the mechanism that has had many experimental measurements for comparison. For the 1PA process, this rule takes the form ${\rm{W}} = 7.43 \times \sqrt {\mathrm{B}} \left( {{\rm{meV;T}}} \right)$. This result agrees well with previous theoretical calculations \cite{bdh,phuc,tuan} and is relatively consistent with experimental observations \cite{ji} and theoretical calculations using the Kubo formula \cite{yang}. For the 2PA process, we find the rule ${\rm{W}} = 3.63 \times \sqrt {\mathrm{B}} \left( {{\rm{meV;T}}} \right)$. This result is also consistent with calculations based on perturbation theory \cite{phuc}. Finally, for the 3PA process, we find the rule ${\rm{W}} = 1.80 \times \sqrt {\mathrm{B}} \left( {{\rm{meV;T}}} \right)$. Next, for the electron-acoustic phonon scattering mechanism, the laws of dependence of HWHM on the magnetic field are shown in Fig. \ref{7b} and Tab. \ref{tab2}. The results show that the HWHM in this scattering process is much smaller than that in the electron-optical phonon scattering process. This is our own new discovery that has not been published before. Unfortunately, for processes that absorb more than one photon in both two cases of the electron scattering mechanism, no experimental measurements have been performed to verify our calculations. We hope that the results of our study will shed some light on future experimental investigations. 

\begin{table}[!htb]
\centering
\caption{The square root law shows the dependence of HWHM on magnetic field (both electron scattering mechanisms) and temperature (electron-acoustic phonon scattering mechanism) in Figs. \ref{7a}, \ref{7b}, and \ref{8b}.}
\begin{ruledtabular}
\begin{tabular}{cccc}
 &
  \begin{tabular}[c]{@{}c@{}}\textbf{Fig. 7(a)}\\ $\rm{W} = \kappa \sqrt{B}$ (meV; T)\end{tabular} &
  \begin{tabular}[c]{@{}c@{}}\textbf{Fig. 7(b)}\\ $\rm{W} = \kappa \sqrt{B}$ (meV; T)\end{tabular} &
  \begin{tabular}[c]{@{}c@{}}\textbf{Fig. 8(b)}\\ $\rm{W} = \kappa \sqrt{T}$ (meV; K)\end{tabular} \\ \hline
\textbf{1PA} & $\rm{W} = 7.43 \sqrt{B}$ & $\rm{W} = 0.059 \sqrt{B}$ & $\rm{W} = 0.064 \sqrt{T}$ \\
\textbf{2PA} & $\rm{W} = 3.63 \sqrt{B}$ & $\rm{W} = 0.029 \sqrt{B}$ & $\rm{W} = 0.032 \sqrt{T}$ \\
\textbf{3PA} & $\rm{W} = 1.80 \sqrt{B}$ & $\rm{W} = 0.020 \sqrt{B}$ & $\rm{W} = 0.022 \sqrt{T}$ \\ 
\end{tabular}
\end{ruledtabular}
\label{tab2}
\end{table}

Temperature is an external parameter that has an important influence on kinetic effects in low-dimensional electron systems. Many previous studies on two-dimensional quantum wells \cite{van3, lvtung, pham} and one-dimensional quantum wires \cite{pro} with different potential profiles have shown that the absorption spectral line width or FWHM is proportional to the square root of the temperature (for the phonon absorption process). For Graphene, we use numerical methods to perform HWHM calculations as a function of temperature and show them in Figs. \ref{8a} and \ref{8b} in both two cases of the electron scattering mechanism. For the electron-optical phonon scattering mechanism (see Fig. \ref{8a}), our calculation results show that HWHM is almost independent of temperature. The HWHM of the 1PA process is larger than the HWHM of the 2PA and 3PA processes. Our HWHM calculations are consistent with experimental observations \cite{orlita} and other theoretical calculations \cite{phuc} for both different values of the external magnetic field for the 1PA process. For the electron-acoustic phonon scattering mechanism, our calculation results show that HWHM increases slightly with temperature as a square root function, shown in Fig. \ref{8b} and Tab. \ref{tab2}. The HWHM in this case is only $10\%$ of the HWHM in the case of electron-optical phonon scattering. This is in contrast to traditional low-dimensional semiconductor systems, where temperature has a strong influence on electron-phonon scattering probability and electron mobility, thereby increasing HWHM \cite{van2, van3}. The temperature range considered in Figs. {\ref{8a}} and {\ref{8b}} correspond to different scattering mechanisms. Specifically, in the high temperature region,  electron-optical phonon scattering is dominant, while in the low temperature region, electron-acoustic phonon scattering is significant. This limit has been used in theoretical calculations in traditional two-dimensional semiconductor systems such as quantum wells {\cite{van1}} and two-dimensional Graphene {\cite{tuan}} allowing to explain the influence of lattice temperature on the transport properties of the carriers. In Graphene, although the contribution of the electron-acoustic phonon scattering process to the value of HWHM is relatively small, the influence of temperature on HWHM cannot be ignored, especially in the low temperature region. To the best of our knowledge, there is currently no quantitative theoretical study showing the influence of the electron-acoustic phonon scattering on HWHM nor any experimental observations to verify our calculations. Therefore, this result is a new finding of ours and can be seen as a useful suggestion for future experimental observations when investigating HWHM to better understand the transport properties of electron in Graphene.

\section{Conclusions}
\begin{figure}[!htb]
\centering
\subfigure[][Electron-optical phonon scattering. Here, ${\mathrm{T}} = 4.2 {\mathrm{K}}, {\mathrm{E}}_{0} = 0.5 \times 10^{7} {\mathrm{V/m}}$.]{\label{7a}\includegraphics[width=0.47\linewidth]{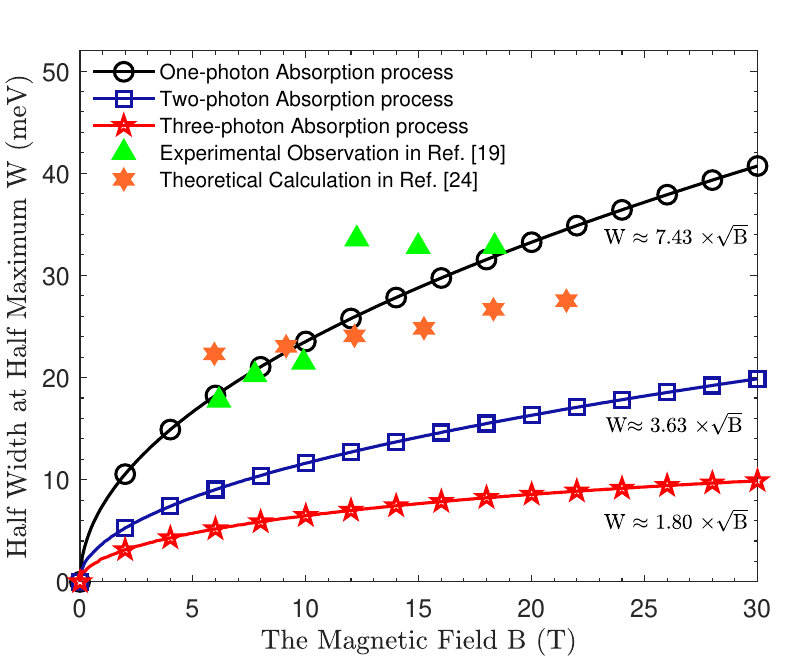}}
\subfigure[][Electron-acoustic phonon scattering. Here, ${\mathrm{T}} = 4.2 {\mathrm{K}}, {\mathrm{E}}_{0} = 2 \times 10^{6} {\mathrm{V/m}}$.]{\label{7b}\includegraphics[width=0.51\linewidth]{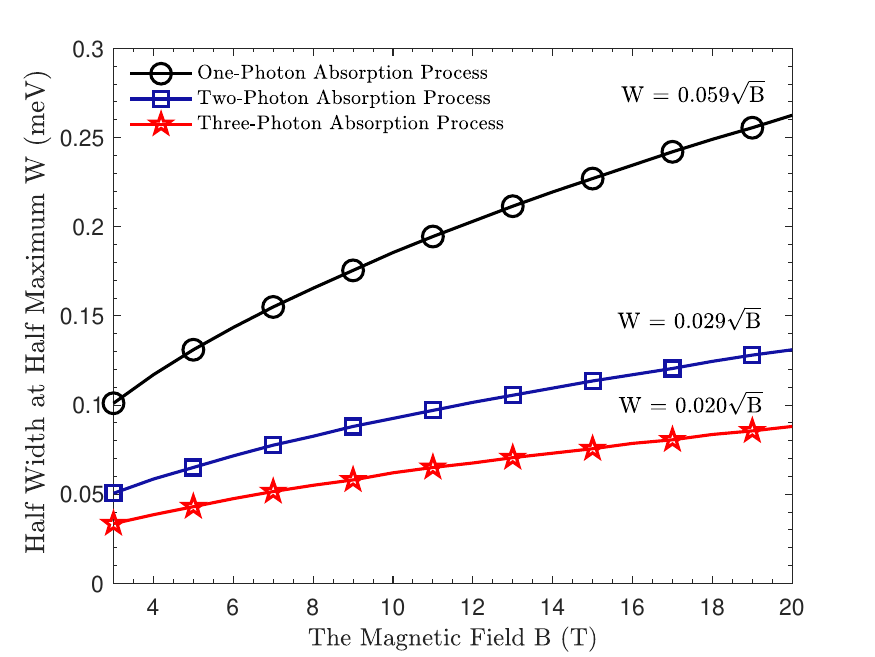}}
\caption{(Color online) HWHM (W) as a function of the magnetic field.}
\label{7}
\end{figure}

\begin{figure}[!htb]
\centering
\subfigure[][Electron-optical phonon scattering.]{\label{8a}\includegraphics[width=0.49\linewidth]{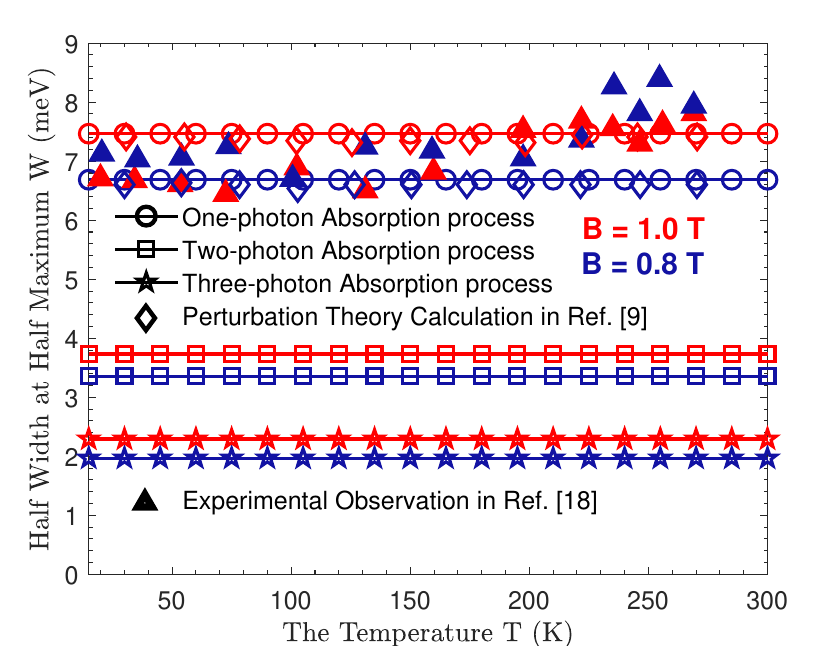}}
\subfigure[][Electron-acoustic phonon scattering. ${\mathrm{B}} = 6 {\mathrm{T}}$]{\label{8b}\includegraphics[width=0.50\linewidth]{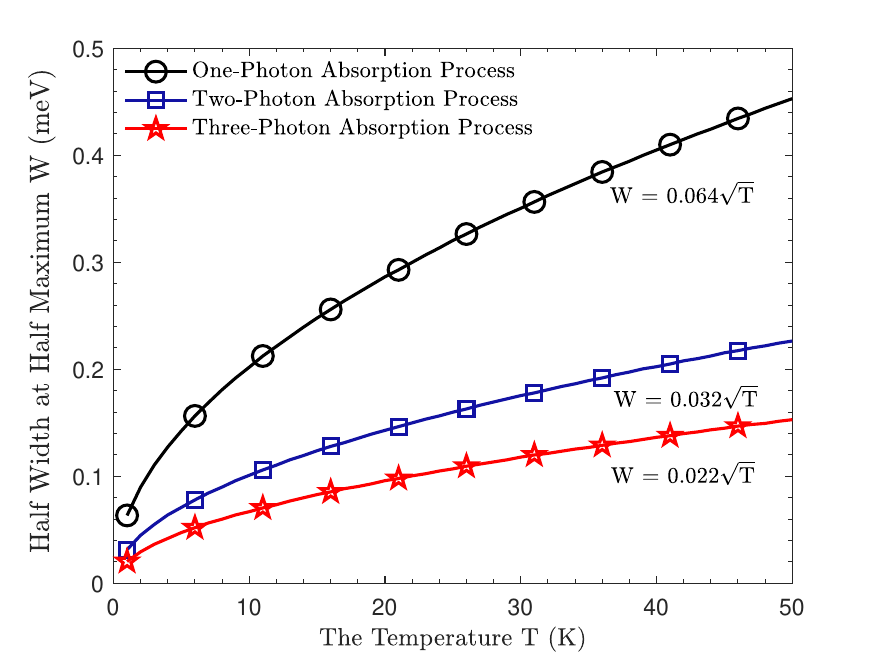}}
\caption{(Color online) HWHM (W) as a function of the temperature of 2DGS. }
\label{8}
\end{figure}

In this study, we have systematically scrutinized the influence of the magnetic field and temperature on the absorption spectrum and HWHM in Graphene in the presence of the SEMW in both electron scattering mechanisms. The main results are summarized as follows. First, we find the MPRC that allows us to detect the positions of the resonance peaks. The greater the external magnetic field and intensity of SEMW, the greater the intensity of the resonance peaks. The intensity of the resonance peak of the 3PA process is $23.3\%$ of the intensity of the 1PA resonance peak when $\rm{E}_0 = 1.5 \times 10^7 \mathrm{V/m}$. Second, the influence of the magnetic field on HWHM is significant, HWHM increases according to the law of proportionality with the square root of the magnetic field as shown in Figs. \ref{7a} and \ref{7b}. In particular, with multi-photon absorption processes, the rules we found are new and valuable in guiding future experimental observations. Finally, the effect of temperature on HWHM in Graphene can be ignored in the case of electron-optical phonon scattering, but cannot be ignored in the case of electron-acoustic phonon scattering (although this effect is quite small). This can be seen as a subtle suggestion for future experimental observations. Our calculations of the dependence of HWHM of the 1PA process on temperature in the electron-optical phonon scattering mechanism are in accordance with previous experimental observations and theoretical calculations.

\begin{acknowledgments}
This research is funded by Vietnam National University, Hanoi - Grant Number QG.22.11. 
\end{acknowledgments}
\bibliography{ref.bib}

\providecommand{\noopsort}[1]{}\providecommand{\singleletter}[1]{#1}%
\begin{thebibliography}{24}%
\makeatletter
\providecommand \@ifxundefined [1]{%
 \@ifx{#1\undefined}
}%
\providecommand \@ifnum [1]{%
 \ifnum #1\expandafter \@firstoftwo
 \else \expandafter \@secondoftwo
 \fi
}%
\providecommand \@ifx [1]{%
 \ifx #1\expandafter \@firstoftwo
 \else \expandafter \@secondoftwo
 \fi
}%
\providecommand \natexlab [1]{#1}%
\providecommand \enquote  [1]{``#1''}%
\providecommand \bibnamefont  [1]{#1}%
\providecommand \bibfnamefont [1]{#1}%
\providecommand \citenamefont [1]{#1}%
\providecommand \href@noop [0]{\@secondoftwo}%
\providecommand \href [0]{\begingroup \@sanitize@url \@href}%
\providecommand \@href[1]{\@@startlink{#1}\@@href}%
\providecommand \@@href[1]{\endgroup#1\@@endlink}%
\providecommand \@sanitize@url [0]{\catcode `\\12\catcode `\$12\catcode `\&12\catcode `\#12\catcode `\^12\catcode `\_12\catcode `\%12\relax}%
\providecommand \@@startlink[1]{}%
\providecommand \@@endlink[0]{}%
\providecommand \url  [0]{\begingroup\@sanitize@url \@url }%
\providecommand \@url [1]{\endgroup\@href {#1}{\urlprefix }}%
\providecommand \urlprefix  [0]{URL }%
\providecommand \Eprint [0]{\href }%
\providecommand \doibase [0]{http://dx.doi.org/}%
\providecommand \selectlanguage [0]{\@gobble}%
\providecommand \bibinfo  [0]{\@secondoftwo}%
\providecommand \bibfield  [0]{\@secondoftwo}%
\providecommand \translation [1]{[#1]}%
\providecommand \BibitemOpen [0]{}%
\providecommand \bibitemStop [0]{}%
\providecommand \bibitemNoStop [0]{.\EOS\space}%
\providecommand \EOS [0]{\spacefactor3000\relax}%
\providecommand \BibitemShut  [1]{\csname bibitem#1\endcsname}%
\let\auto@bib@innerbib\@empty
\bibitem [{\citenamefont {Novoselov}\ \emph {et~al.}(2004)\citenamefont {Novoselov}, \citenamefont {Geim}, \citenamefont {Morozov}, \citenamefont {Jiang}, \citenamefont {Zhang}, \citenamefont {Dubonos}, \citenamefont {Grigorieva},\ and\ \citenamefont {Firsov}}]{akgeim}%
  \BibitemOpen
  \bibfield  {author} {\bibinfo {author} {\bibfnamefont {K.~S.}\ \bibnamefont {Novoselov}}, \bibinfo {author} {\bibfnamefont {A.~K.}\ \bibnamefont {Geim}}, \bibinfo {author} {\bibfnamefont {S.~V.}\ \bibnamefont {Morozov}}, \bibinfo {author} {\bibfnamefont {D.}~\bibnamefont {Jiang}}, \bibinfo {author} {\bibfnamefont {Y.}~\bibnamefont {Zhang}}, \bibinfo {author} {\bibfnamefont {S.~V.}\ \bibnamefont {Dubonos}}, \bibinfo {author} {\bibfnamefont {I.~V.}\ \bibnamefont {Grigorieva}}, \ and\ \bibinfo {author} {\bibfnamefont {A.~A.}\ \bibnamefont {Firsov}},\ }\href@noop {} {\bibfield  {journal} {\bibinfo  {journal} {Science}\ }\textbf {\bibinfo {volume} {306}},\ \bibinfo {pages} {666} (\bibinfo {year} {2004})}\BibitemShut {NoStop}%
\bibitem [{\citenamefont {Novoselov}\ \emph {et~al.}(2005)\citenamefont {Novoselov}, \citenamefont {Geim}, \citenamefont {Morozov}, \citenamefont {Jiang}, \citenamefont {Katsnelson}, \citenamefont {Grigorieva}, \citenamefont {Dubonos},\ and\ \citenamefont {Firsov}}]{ksno}%
  \BibitemOpen
  \bibfield  {author} {\bibinfo {author} {\bibfnamefont {K.~S.}\ \bibnamefont {Novoselov}}, \bibinfo {author} {\bibfnamefont {A.~K.}\ \bibnamefont {Geim}}, \bibinfo {author} {\bibfnamefont {S.~V.}\ \bibnamefont {Morozov}}, \bibinfo {author} {\bibfnamefont {D.}~\bibnamefont {Jiang}}, \bibinfo {author} {\bibfnamefont {M.~I.}\ \bibnamefont {Katsnelson}}, \bibinfo {author} {\bibfnamefont {I.~V.}\ \bibnamefont {Grigorieva}}, \bibinfo {author} {\bibfnamefont {S.~V.}\ \bibnamefont {Dubonos}}, \ and\ \bibinfo {author} {\bibfnamefont {A.~A.}\ \bibnamefont {Firsov}},\ }\href@noop {} {\bibfield  {journal} {\bibinfo  {journal} {Nature.}\ }\textbf {\bibinfo {volume} {438}},\ \bibinfo {pages} {197200} (\bibinfo {year} {2005})}\BibitemShut {NoStop}%
\bibitem [{\citenamefont {Malevich}\ and\ \citenamefont {Epstein}(1974)}]{eps}%
  \BibitemOpen
  \bibfield  {author} {\bibinfo {author} {\bibfnamefont {V.~L.}\ \bibnamefont {Malevich}}\ and\ \bibinfo {author} {\bibfnamefont {E.~M.}\ \bibnamefont {Epstein}},\ }\href@noop {} {\bibfield  {journal} {\bibinfo  {journal} {Sov. Quantum Electronic.}\ }\textbf {\bibinfo {volume} {1}},\ \bibinfo {pages} {1468} (\bibinfo {year} {1974})}\BibitemShut {NoStop}%
\bibitem [{\citenamefont {Pavlovich}\ and\ \citenamefont {Epstein}(1977)}]{eps1}%
  \BibitemOpen
  \bibfield  {author} {\bibinfo {author} {\bibfnamefont {V.~V.}\ \bibnamefont {Pavlovich}}\ and\ \bibinfo {author} {\bibfnamefont {E.~M.}\ \bibnamefont {Epstein}},\ }\href@noop {} {\bibfield  {journal} {\bibinfo  {journal} {Sov. Phys. Stat.}\ }\textbf {\bibinfo {volume} {19}},\ \bibinfo {pages} {1760} (\bibinfo {year} {1977})}\BibitemShut {NoStop}%
\bibitem [{\citenamefont {Bau}\ and\ \citenamefont {Phong}(1998)}]{bau2}%
  \BibitemOpen
  \bibfield  {author} {\bibinfo {author} {\bibfnamefont {N.~Q.}\ \bibnamefont {Bau}}\ and\ \bibinfo {author} {\bibfnamefont {T.~C.}\ \bibnamefont {Phong}},\ }\href@noop {} {\bibfield  {journal} {\bibinfo  {journal} {J. Phys. Soc. Jpn.}\ }\textbf {\bibinfo {volume} {67}},\ \bibinfo {pages} {3875} (\bibinfo {year} {1998})}\BibitemShut {NoStop}%
\bibitem [{\citenamefont {Hoi}\ \emph {et~al.}(2018)\citenamefont {Hoi}, \citenamefont {Phuong},\ and\ \citenamefont {Phong}}]{bdh}%
  \BibitemOpen
  \bibfield  {author} {\bibinfo {author} {\bibfnamefont {B.~D.}\ \bibnamefont {Hoi}}, \bibinfo {author} {\bibfnamefont {L.~T.~T.}\ \bibnamefont {Phuong}}, \ and\ \bibinfo {author} {\bibfnamefont {T.~C.}\ \bibnamefont {Phong}},\ }\href@noop {} {\bibfield  {journal} {\bibinfo  {journal} {J. Appl. Phys.}\ }\textbf {\bibinfo {volume} {123}},\ \bibinfo {pages} {094303} (\bibinfo {year} {2018})}\BibitemShut {NoStop}%
\bibitem [{\citenamefont {Tung}\ \emph {et~al.}(2019)\citenamefont {Tung}, \citenamefont {Lam}, \citenamefont {Bau}, \citenamefont {Huyen}, \citenamefont {Phuc},\ and\ \citenamefont {Nguyen}}]{lvtung}%
  \BibitemOpen
  \bibfield  {author} {\bibinfo {author} {\bibfnamefont {L.~V.}\ \bibnamefont {Tung}}, \bibinfo {author} {\bibfnamefont {V.~T.}\ \bibnamefont {Lam}}, \bibinfo {author} {\bibfnamefont {N.~Q.}\ \bibnamefont {Bau}}, \bibinfo {author} {\bibfnamefont {P.~T.~K.}\ \bibnamefont {Huyen}}, \bibinfo {author} {\bibfnamefont {H.~V.}\ \bibnamefont {Phuc}}, \ and\ \bibinfo {author} {\bibfnamefont {C.~V.}\ \bibnamefont {Nguyen}},\ }\href@noop {} {\bibfield  {journal} {\bibinfo  {journal} {Superlattice Microst.}\ }\textbf {\bibinfo {volume} {130}},\ \bibinfo {pages} {446453} (\bibinfo {year} {2019})}\BibitemShut {NoStop}%
\bibitem [{\citenamefont {Pham}\ \emph {et~al.}(2019)\citenamefont {Pham}, \citenamefont {Tung}, \citenamefont {Thuan}, \citenamefont {Nguyen}, \citenamefont {Hieu},\ and\ \citenamefont {Phuc}}]{pham}%
  \BibitemOpen
  \bibfield  {author} {\bibinfo {author} {\bibfnamefont {K.~D.}\ \bibnamefont {Pham}}, \bibinfo {author} {\bibfnamefont {L.~V.}\ \bibnamefont {Tung}}, \bibinfo {author} {\bibfnamefont {D.~V.}\ \bibnamefont {Thuan}}, \bibinfo {author} {\bibfnamefont {C.~V.}\ \bibnamefont {Nguyen}}, \bibinfo {author} {\bibfnamefont {N.~N.}\ \bibnamefont {Hieu}}, \ and\ \bibinfo {author} {\bibfnamefont {H.~V.}\ \bibnamefont {Phuc}},\ }\href@noop {} {\bibfield  {journal} {\bibinfo  {journal} {J. Appl. Phys.}\ }\textbf {\bibinfo {volume} {126}} (\bibinfo {year} {2019})}\BibitemShut {NoStop}%
\bibitem [{\citenamefont {Phuc}\ and\ \citenamefont {Hieu}(2015)}]{phuc}%
  \BibitemOpen
  \bibfield  {author} {\bibinfo {author} {\bibfnamefont {H.~V.}\ \bibnamefont {Phuc}}\ and\ \bibinfo {author} {\bibfnamefont {N.~N.}\ \bibnamefont {Hieu}},\ }\href@noop {} {\bibfield  {journal} {\bibinfo  {journal} {Opt. Commun.}\ }\textbf {\bibinfo {volume} {344}},\ \bibinfo {pages} {1216} (\bibinfo {year} {2015})}\BibitemShut {NoStop}%
\bibitem [{\citenamefont {Dakhlaoui}(2015)}]{hjap}%
  \BibitemOpen
  \bibfield  {author} {\bibinfo {author} {\bibfnamefont {H.}~\bibnamefont {Dakhlaoui}},\ }\href@noop {} {\bibfield  {journal} {\bibinfo  {journal} {J. Appl. Phys.}\ }\textbf {\bibinfo {volume} {117}},\ \bibinfo {pages} {135705} (\bibinfo {year} {2015})}\BibitemShut {NoStop}%
\bibitem [{\citenamefont {Kryuchkov}\ \emph {et~al.}(2013)\citenamefont {Kryuchkov}, \citenamefont {Kukhar},\ and\ \citenamefont {Zavyalov}}]{kry}%
  \BibitemOpen
  \bibfield  {author} {\bibinfo {author} {\bibfnamefont {S.~V.}\ \bibnamefont {Kryuchkov}}, \bibinfo {author} {\bibfnamefont {E.~I.}\ \bibnamefont {Kukhar}}, \ and\ \bibinfo {author} {\bibfnamefont {D.~V.}\ \bibnamefont {Zavyalov}},\ }\href@noop {} {\bibfield  {journal} {\bibinfo  {journal} {Phys. Wave Phenom.}\ }\textbf {\bibinfo {volume} {21}},\ \bibinfo {pages} {207213} (\bibinfo {year} {2013})}\BibitemShut {NoStop}%
\bibitem [{\citenamefont {Bau}\ \emph {et~al.}(2009)\citenamefont {Bau}, \citenamefont {Hung},\ and\ \citenamefont {Ngoc}}]{bau1}%
  \BibitemOpen
  \bibfield  {author} {\bibinfo {author} {\bibfnamefont {N.~Q.}\ \bibnamefont {Bau}}, \bibinfo {author} {\bibfnamefont {D.~M.}\ \bibnamefont {Hung}}, \ and\ \bibinfo {author} {\bibfnamefont {N.~B.}\ \bibnamefont {Ngoc}},\ }\href@noop {} {\bibfield  {journal} {\bibinfo  {journal} {J. Korean Phys. Soc.}\ }\textbf {\bibinfo {volume} {54}},\ \bibinfo {pages} {765773} (\bibinfo {year} {2009})}\BibitemShut {NoStop}%
\bibitem [{\citenamefont {Tuan}\ \emph {et~al.}(2023)\citenamefont {Tuan}, \citenamefont {Bau}, \citenamefont {Nam}, \citenamefont {Ba},\ and\ \citenamefont {Nhan}}]{tuan}%
  \BibitemOpen
  \bibfield  {author} {\bibinfo {author} {\bibfnamefont {T.~A.}\ \bibnamefont {Tuan}}, \bibinfo {author} {\bibfnamefont {N.~Q.}\ \bibnamefont {Bau}}, \bibinfo {author} {\bibfnamefont {N.~D.}\ \bibnamefont {Nam}}, \bibinfo {author} {\bibfnamefont {C.~T.~V.}\ \bibnamefont {Ba}}, \ and\ \bibinfo {author} {\bibfnamefont {N.~T.~T.}\ \bibnamefont {Nhan}},\ }\href@noop {} {\bibfield  {journal} {\bibinfo  {journal} {J. Phys. Soc. Jpn.}\ }\textbf {\bibinfo {volume} {92}},\ \bibinfo {pages} {064401} (\bibinfo {year} {2023})}\BibitemShut {NoStop}%
\bibitem [{\citenamefont {Unuma}\ \emph {et~al.}(2003)\citenamefont {Unuma}, \citenamefont {Yoshita}, \citenamefont {Noda}, \citenamefont {Sakaki},\ and\ \citenamefont {Akiyama}}]{unuma}%
  \BibitemOpen
  \bibfield  {author} {\bibinfo {author} {\bibfnamefont {T.}~\bibnamefont {Unuma}}, \bibinfo {author} {\bibfnamefont {M.}~\bibnamefont {Yoshita}}, \bibinfo {author} {\bibfnamefont {T.}~\bibnamefont {Noda}}, \bibinfo {author} {\bibfnamefont {H.}~\bibnamefont {Sakaki}}, \ and\ \bibinfo {author} {\bibfnamefont {H.}~\bibnamefont {Akiyama}},\ }\href@noop {} {\bibfield  {journal} {\bibinfo  {journal} {J. Appl. Phys.}\ }\textbf {\bibinfo {volume} {93}},\ \bibinfo {pages} {1586} (\bibinfo {year} {2003})}\BibitemShut {NoStop}%
\bibitem [{\citenamefont {Campman}\ \emph {et~al.}(1996)\citenamefont {Campman}, \citenamefont {Schmidt}, \citenamefont {Imamoglu},\ and\ \citenamefont {Gossard}}]{tndr1}%
  \BibitemOpen
  \bibfield  {author} {\bibinfo {author} {\bibfnamefont {K.}~\bibnamefont {Campman}}, \bibinfo {author} {\bibfnamefont {H.}~\bibnamefont {Schmidt}}, \bibinfo {author} {\bibfnamefont {A.}~\bibnamefont {Imamoglu}}, \ and\ \bibinfo {author} {\bibfnamefont {A.}~\bibnamefont {Gossard}},\ }\href@noop {} {\bibfield  {journal} {\bibinfo  {journal} {Appl. Phys. Lett.}\ }\textbf {\bibinfo {volume} {69}},\ \bibinfo {pages} {2554} (\bibinfo {year} {1996})}\BibitemShut {NoStop}%
\bibitem [{\citenamefont {Dupont}\ \emph {et~al.}(1992)\citenamefont {Dupont}, \citenamefont {Delacourt}, \citenamefont {Papillon}, \citenamefont {Schnell},\ and\ \citenamefont {Papuchon}}]{tndr2}%
  \BibitemOpen
  \bibfield  {author} {\bibinfo {author} {\bibfnamefont {E.}~\bibnamefont {Dupont}}, \bibinfo {author} {\bibfnamefont {D.}~\bibnamefont {Delacourt}}, \bibinfo {author} {\bibfnamefont {D.}~\bibnamefont {Papillon}}, \bibinfo {author} {\bibfnamefont {J.}~\bibnamefont {Schnell}}, \ and\ \bibinfo {author} {\bibfnamefont {M.}~\bibnamefont {Papuchon}},\ }\href@noop {} {\bibfield  {journal} {\bibinfo  {journal} {Appl. Phys. Lett.}\ }\textbf {\bibinfo {volume} {60}},\ \bibinfo {pages} {2121} (\bibinfo {year} {1992})}\BibitemShut {NoStop}%
\bibitem [{\citenamefont {Chaubey}\ and\ \citenamefont {Van~Vliet}(1986{\natexlab{a}})}]{van2}%
  \BibitemOpen
  \bibfield  {author} {\bibinfo {author} {\bibfnamefont {M.}~\bibnamefont {Chaubey}}\ and\ \bibinfo {author} {\bibfnamefont {C.}~\bibnamefont {Van~Vliet}},\ }\href@noop {} {\bibfield  {journal} {\bibinfo  {journal} {Phys. Rev. B.}\ }\textbf {\bibinfo {volume} {34}},\ \bibinfo {pages} {3932} (\bibinfo {year} {1986}{\natexlab{a}})}\BibitemShut {NoStop}%
\bibitem [{\citenamefont {Orlita}\ \emph {et~al.}(2008)\citenamefont {Orlita}, \citenamefont {Faugeras}, \citenamefont {Plochocka}, \citenamefont {Neugebauer}, \citenamefont {Martinez}, \citenamefont {Maude}, \citenamefont {Barra}, \citenamefont {Sprinkle}, \citenamefont {Berger}, \citenamefont {de~Heer},\ and\ \citenamefont {Potemski}}]{orlita}%
  \BibitemOpen
  \bibfield  {author} {\bibinfo {author} {\bibfnamefont {M.}~\bibnamefont {Orlita}}, \bibinfo {author} {\bibfnamefont {C.}~\bibnamefont {Faugeras}}, \bibinfo {author} {\bibfnamefont {P.}~\bibnamefont {Plochocka}}, \bibinfo {author} {\bibfnamefont {P.}~\bibnamefont {Neugebauer}}, \bibinfo {author} {\bibfnamefont {G.}~\bibnamefont {Martinez}}, \bibinfo {author} {\bibfnamefont {D.~K.}\ \bibnamefont {Maude}}, \bibinfo {author} {\bibfnamefont {A.-L.}\ \bibnamefont {Barra}}, \bibinfo {author} {\bibfnamefont {M.}~\bibnamefont {Sprinkle}}, \bibinfo {author} {\bibfnamefont {C.}~\bibnamefont {Berger}}, \bibinfo {author} {\bibfnamefont {W.~A.}\ \bibnamefont {de~Heer}}, \ and\ \bibinfo {author} {\bibfnamefont {M.}~\bibnamefont {Potemski}},\ }\href@noop {} {\bibfield  {journal} {\bibinfo  {journal} {Phys. Rev. Lett.}\ }\textbf {\bibinfo {volume} {101}},\ \bibinfo {pages} {267601} (\bibinfo {year} {2008})}\BibitemShut {NoStop}%
\bibitem [{\citenamefont {Jiang}\ \emph {et~al.}(2007)\citenamefont {Jiang}, \citenamefont {Henriksen}, \citenamefont {Tung}, \citenamefont {Wang}, \citenamefont {Schwartz}, \citenamefont {Han}, \citenamefont {Kim},\ and\ \citenamefont {Stormer}}]{ji}%
  \BibitemOpen
  \bibfield  {author} {\bibinfo {author} {\bibfnamefont {Z.}~\bibnamefont {Jiang}}, \bibinfo {author} {\bibfnamefont {E.~A.}\ \bibnamefont {Henriksen}}, \bibinfo {author} {\bibfnamefont {L.~C.}\ \bibnamefont {Tung}}, \bibinfo {author} {\bibfnamefont {J.}~\bibnamefont {Wang}}, \bibinfo {author} {\bibfnamefont {M.~E.}\ \bibnamefont {Schwartz}}, \bibinfo {author} {\bibfnamefont {M.~Y.}\ \bibnamefont {Han}}, \bibinfo {author} {\bibfnamefont {P.}~\bibnamefont {Kim}}, \ and\ \bibinfo {author} {\bibfnamefont {H.~L.}\ \bibnamefont {Stormer}},\ }\href@noop {} {\bibfield  {journal} {\bibinfo  {journal} {Phys. Rev. Lett.}\ }\textbf {\bibinfo {volume} {98}},\ \bibinfo {pages} {197403} (\bibinfo {year} {2007})}\BibitemShut {NoStop}%
\bibitem [{\citenamefont {Phong}\ and\ \citenamefont {Phuc}(2011)}]{pro}%
  \BibitemOpen
  \bibfield  {author} {\bibinfo {author} {\bibfnamefont {T.~C.}\ \bibnamefont {Phong}}\ and\ \bibinfo {author} {\bibfnamefont {H.~V.}\ \bibnamefont {Phuc}},\ }\href@noop {} {\bibfield  {journal} {\bibinfo  {journal} {Mod. Phys. Lett. B.}\ }\textbf {\bibinfo {volume} {25}},\ \bibinfo {pages} {1003} (\bibinfo {year} {2011})}\BibitemShut {NoStop}%
\bibitem [{\citenamefont {Mori}\ and\ \citenamefont {Ando}(2011)}]{ando}%
  \BibitemOpen
  \bibfield  {author} {\bibinfo {author} {\bibfnamefont {N.}~\bibnamefont {Mori}}\ and\ \bibinfo {author} {\bibfnamefont {T.}~\bibnamefont {Ando}},\ }\href@noop {} {\bibfield  {journal} {\bibinfo  {journal} {J. Phys. Soc. Jpn.}\ }\textbf {\bibinfo {volume} {80}},\ \bibinfo {pages} {044706} (\bibinfo {year} {2011})}\BibitemShut {NoStop}%
\bibitem [{\citenamefont {Vasilopoulos}(1986)}]{van1}%
  \BibitemOpen
  \bibfield  {author} {\bibinfo {author} {\bibfnamefont {P.}~\bibnamefont {Vasilopoulos}},\ }\href@noop {} {\bibfield  {journal} {\bibinfo  {journal} {Phys. Rev. B.}\ }\textbf {\bibinfo {volume} {33}},\ \bibinfo {pages} {8587} (\bibinfo {year} {1986})}\BibitemShut {NoStop}%
\bibitem [{\citenamefont {Chaubey}\ and\ \citenamefont {Van~Vliet}(1986{\natexlab{b}})}]{van3}%
  \BibitemOpen
  \bibfield  {author} {\bibinfo {author} {\bibfnamefont {M.~P.}\ \bibnamefont {Chaubey}}\ and\ \bibinfo {author} {\bibfnamefont {C.~M.}\ \bibnamefont {Van~Vliet}},\ }\href@noop {} {\bibfield  {journal} {\bibinfo  {journal} {Phys. Rev. B.}\ }\textbf {\bibinfo {volume} {33}},\ \bibinfo {pages} {5617} (\bibinfo {year} {1986}{\natexlab{b}})}\BibitemShut {NoStop}%
\bibitem [{\citenamefont {Yang}\ \emph {et~al.}(2010)\citenamefont {Yang}, \citenamefont {Peeters},\ and\ \citenamefont {Xu}}]{yang}%
  \BibitemOpen
  \bibfield  {author} {\bibinfo {author} {\bibfnamefont {C.}~\bibnamefont {Yang}}, \bibinfo {author} {\bibfnamefont {F.}~\bibnamefont {Peeters}}, \ and\ \bibinfo {author} {\bibfnamefont {W.}~\bibnamefont {Xu}},\ }\href@noop {} {\bibfield  {journal} {\bibinfo  {journal} {Phys. Rev. B.}\ }\textbf {\bibinfo {volume} {82}},\ \bibinfo {pages} {205428} (\bibinfo {year} {2010})}\BibitemShut {NoStop}%
\end{thebibliography}%
\newpage
\end{document}